\documentstyle[editedvolume,numreferences]{crckapb} 

\begin{opening}
\title{REMARKS ON DOMAIN-WALL FERMIONS}

\author{Michael Creutz}
\institute{Physics Department\\
Brookhaven National Laboratory\\
Upton, NY 11973\\
USA\\
creutz@bnl.gov
}

\end{opening}

\begin{document}


\input epsf

\section{Introduction}

Issues of chiral symmetry permeate theoretical physics.  Our
understanding of pionic interactions revolves around spontaneous
symmetry breaking and approximately conserved axial currents.  The
standard model itself is truly chiral, with the weak gauge bosons only
coupling to one helicity state of the fundamental fermions.  In the
context of unification, chiral symmetry provides a mechanism for
protecting fermion masses, possibly explaining how a theory at a much
higher scale can avoid large renormalizations of the light particle
masses.  Extending this mechanism to bosons provides one of the more
compelling motivations for super-symmetry.

On the lattice, chiral symmetry raises many interesting issues.  These
are intricately entwined with the famous axial anomalies and the so
called ``doubling'' problem.  Being a full regulator, the lattice must
break some aspects of chiral symmetry to give the required anomalies
in the continuum limit.  Prescriptions for lattice fermions that do
not accommodate anomalies cancel them with spurious extra species
(doublers).  Domain-wall fermions\cite{dwf}, the motivation for this
talk and the subject of most of todays presentations, are one scheme
to minimize these necessary symmetry violations.

But speak to an audience that already knows all this.  In an attempt
to avoid boring you, I will discuss domain-wall fermions from a rather
unconventional direction.  Following a recent paper of mine
\cite{icetray}, I present the subject from a ``chemists'' point of
view, in terms of a chain molecule with special electronic states
carrying energies fixed by symmetries.  For lattice gauge theory,
placing one of these molecules at each space-time site gives
excitations of naturally zero mass.  This is in direct analogy to the
role of chiral symmetry in conventional continuum descriptions.  After
presenting this picture, I will wander into some comments and
speculations about exact lattice chiral symmetries and schemes for
gauging them.

\section{A ladder molecule}

To start, let me consider two rows of atoms connected by horizontal
and diagonal bonds, as illustrated here

\medskip
\epsfxsize .7\hsize
\centerline {\epsfbox{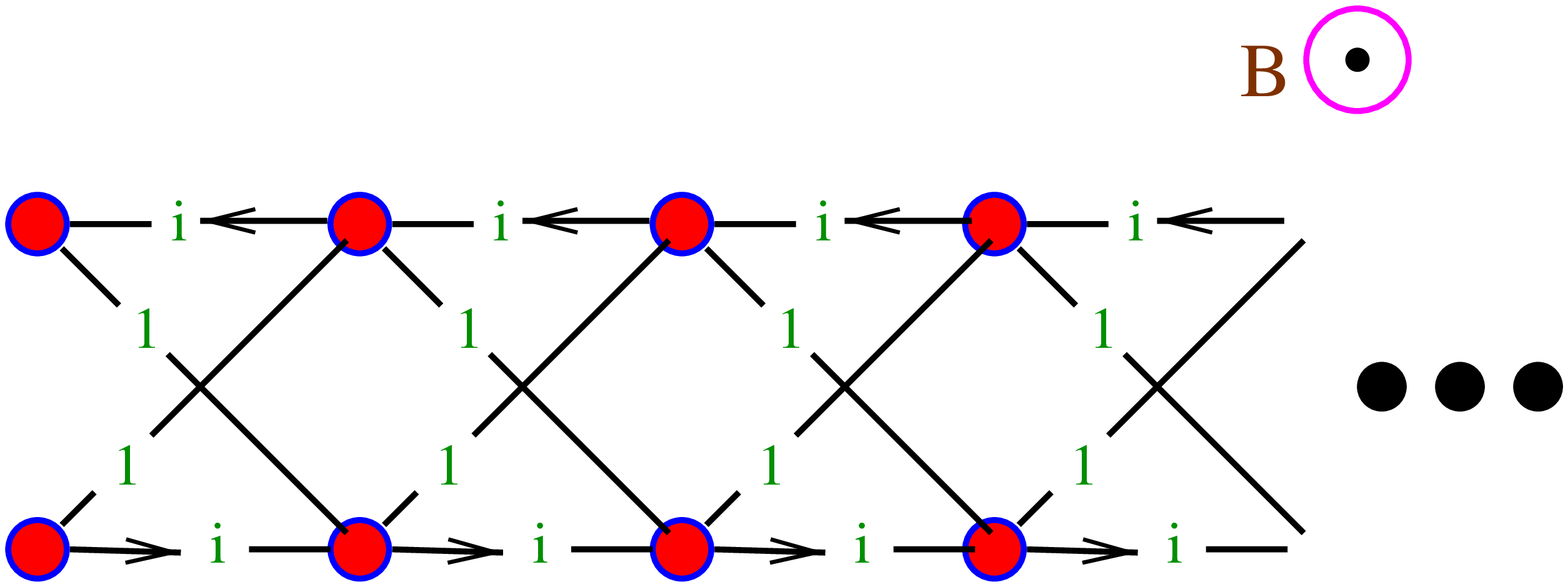}}
\medskip

\noindent The bonds represent hopping terms, wherein an electron moves
from one site to another via a creation-annihilation operator pair in
the Hamiltonian.  Later I will include vertical bonds, but for now
consider just the horizontal and diagonal connections.

Years ago during a course on quantum mechanics, I heard Feynman
present an amusing description of an electron's behavior when inserted
into a lattice.  If you place it initially on a single atom, the wave
function will gradually spread through the lattice, much like water
poured in a cell of a metal ice cube tray.  With damping, it settles
into the ground state which has equal amplitude on each atom.  To this
day I cannot fill an ice cube tray without thinking of this analogy
and pouring all the incoming water into a single cell.

I now complicate this picture with a magnetic field applied
orthogonal to the plane of the system.  This introduces phases as the
electron hops, causing interesting interference effects.  In
particular, consider a field of one-half flux unit per plaquette.
This means that when a particle hops around a unit area (in terms of
the basic lattice spacing) the wave function picks up a minus sign.
Just where the phases appear is a gauge dependent convention; only the
total phase around a closed loop is physical.  One choice for these
phases is indicated by the numbers on the bonds in the above picture.

The phase factors cause cancellations and slow diffusion.  For
example, consider the two shortest paths between the sites {\bf a} and
{\bf b} in the following picture

\medskip
\epsfxsize .4\hsize
\centerline{\epsfbox{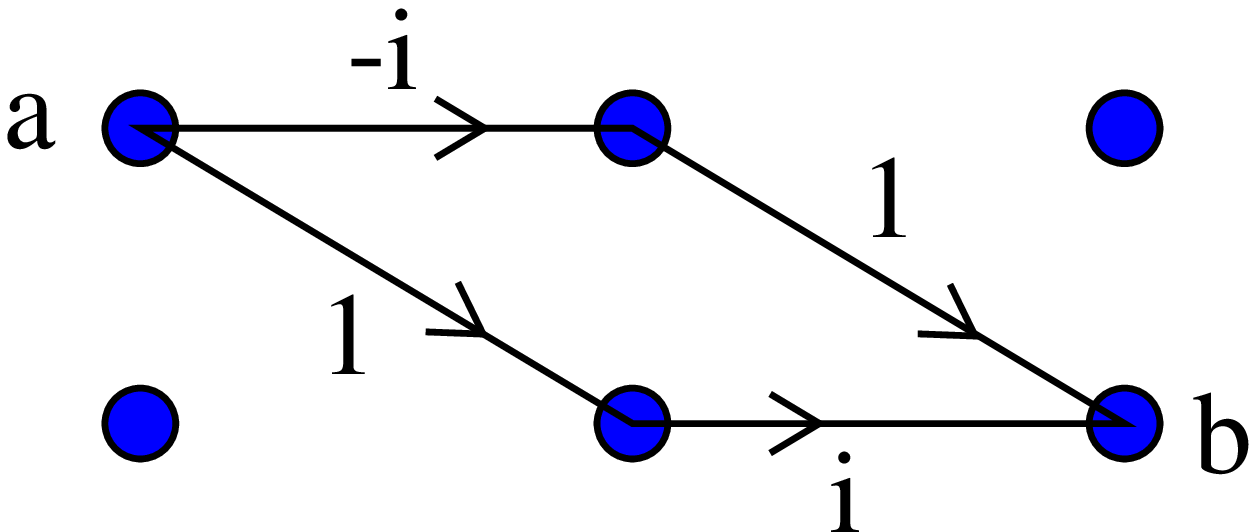}}
\medskip

\noindent With the chosen flux, these paths exactly cancel.  For the
full molecule this cancellation extends to all paths between these
sites.  An electron placed on site {\bf a} can never diffuse to site
{\bf b}.  Unlike in the ice tray analogy, the wave function will not
spread to any site beyond the five nearest neighbors.

As a consequence, the Hamiltonian has localized eigenstates.  While it
is perhaps a bit of a misuse of the term, these states are
``soliton-like'' in that they just sit there and do not change their
shape.  There are two such states per plaquette; one possible
representation for these two states is shown in the following figure

\medskip
\epsfxsize .65\hsize
\centerline {\epsfbox{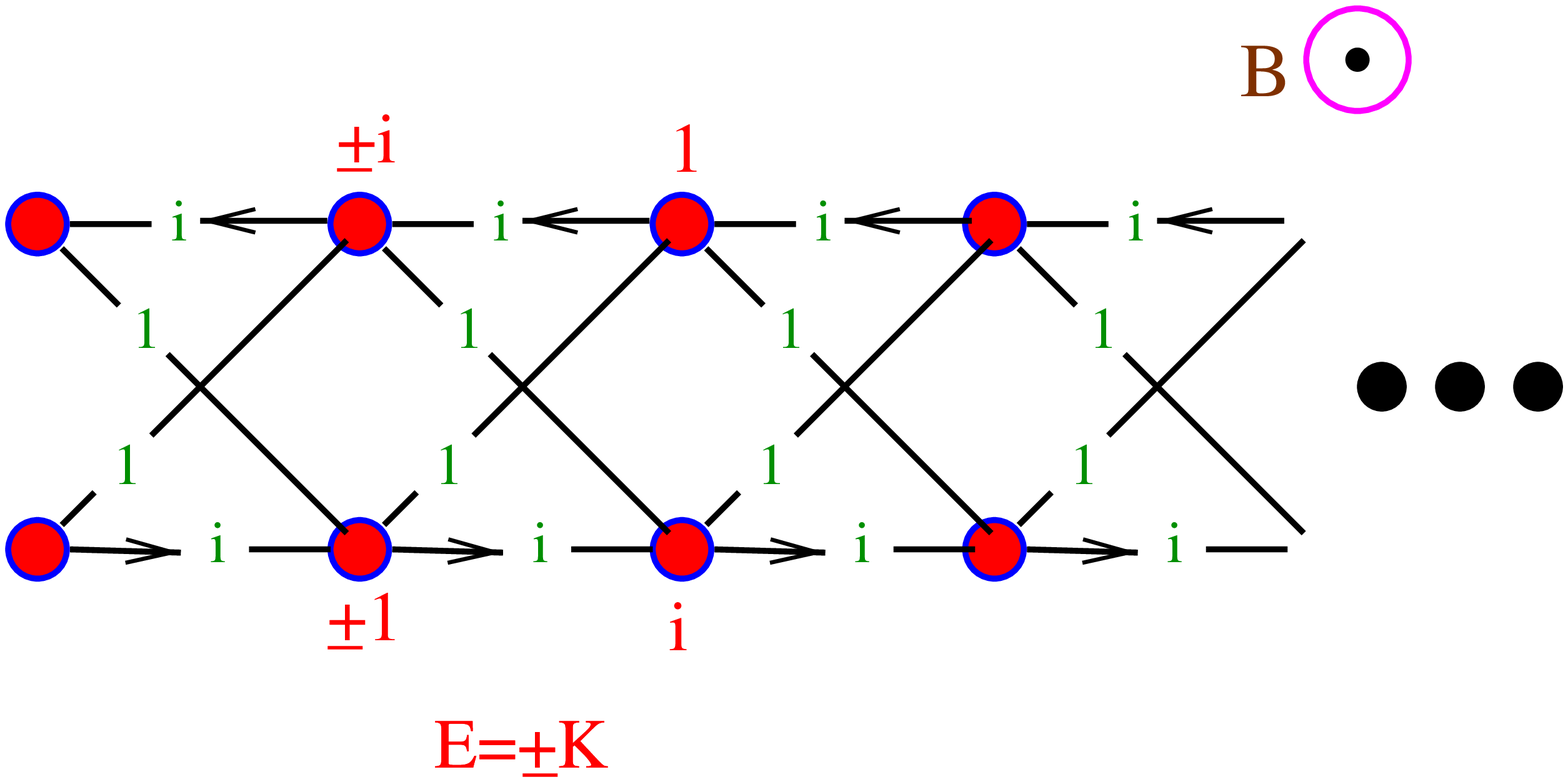}}
\medskip
\noindent The states are restricted to the four sights labeled by
their relative wave functions.  Their energies are fixed by the size
of the hopping parameter $K$.

For a finite chain of length $L$ there are $2L$ atoms, and thus there
should be a total of $2L$ possible states for our electron (ignoring
spin for the moment).  There are $L-1$ plaquettes, and thus $2L-2$ of
the above soliton states.  This is almost the entire spectrum of the
Hamiltonian, but two states are left over.  These are zero energy
states bound to the ends of the system.  The wave function for one of
those is shown here

\medskip
\epsfxsize .65\hsize
\centerline {\epsfbox{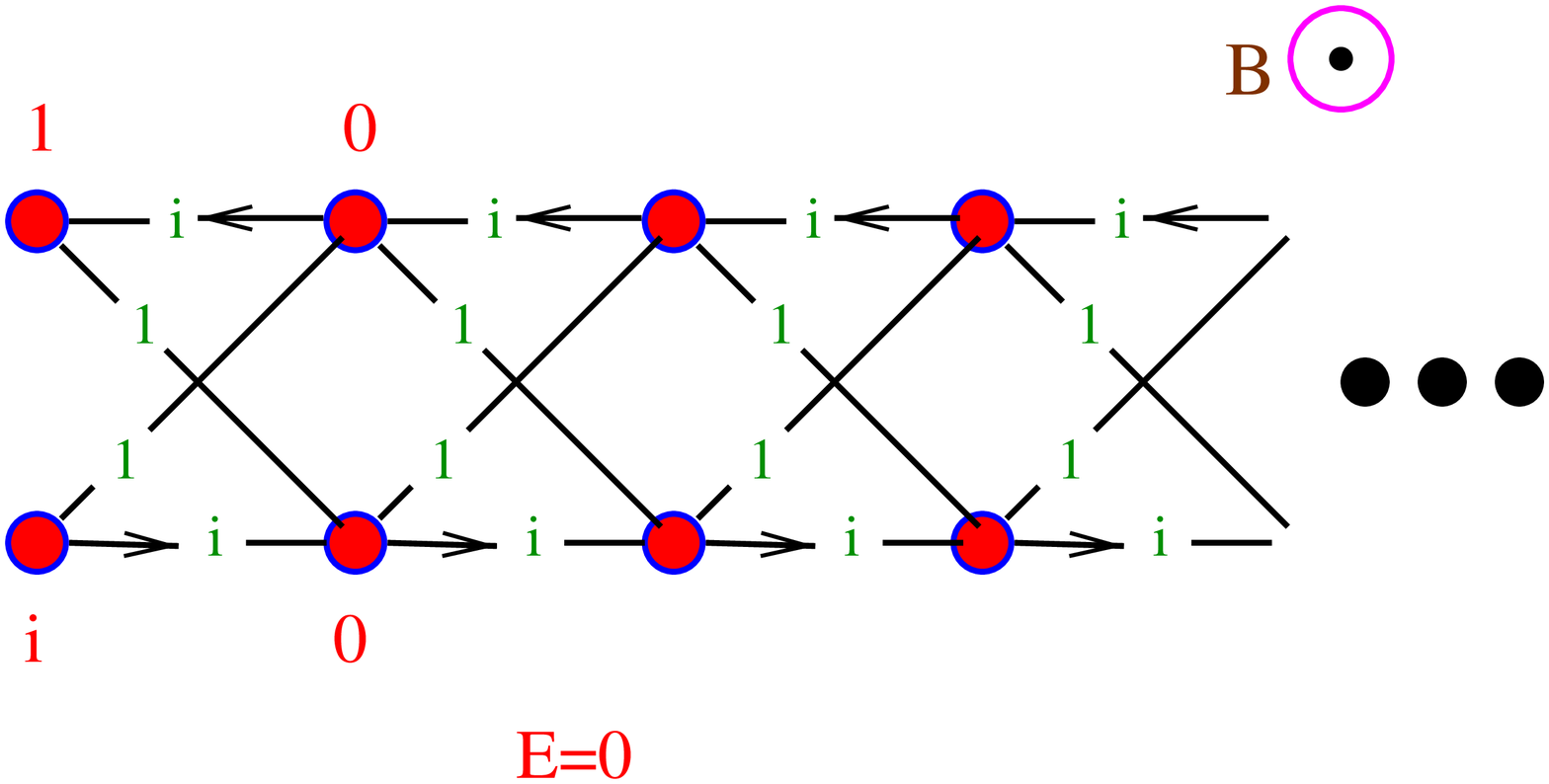}}
\medskip
\noindent We now have the full spectrum of the Hamiltonian: $L-1$
degenerate states of positive energy, a similar number of degenerate
negative energy states, and two states of zero energy bound on the
ends.

Now consider what happens when vertical bonds are included in our
molecule.  The phase cancellations are no longer complete and the
solitonic states spread to form two bands, one with positive and one
with negative energy.  However, for our purposes, the remarkable
result is that the zero modes bound on the ends of the chain are
robust.  The corresponding wave functions are no longer exactly
located on the last atomic pair, but now have an exponentially
suppressed penetration into the chain.  The following figure shows the
wave function for one of these states when the vertical bond has the
same strength as the others.  There is a corresponding state on the
other end of the molecule.

\medskip
\epsfxsize .65\hsize
\centerline {\epsfbox{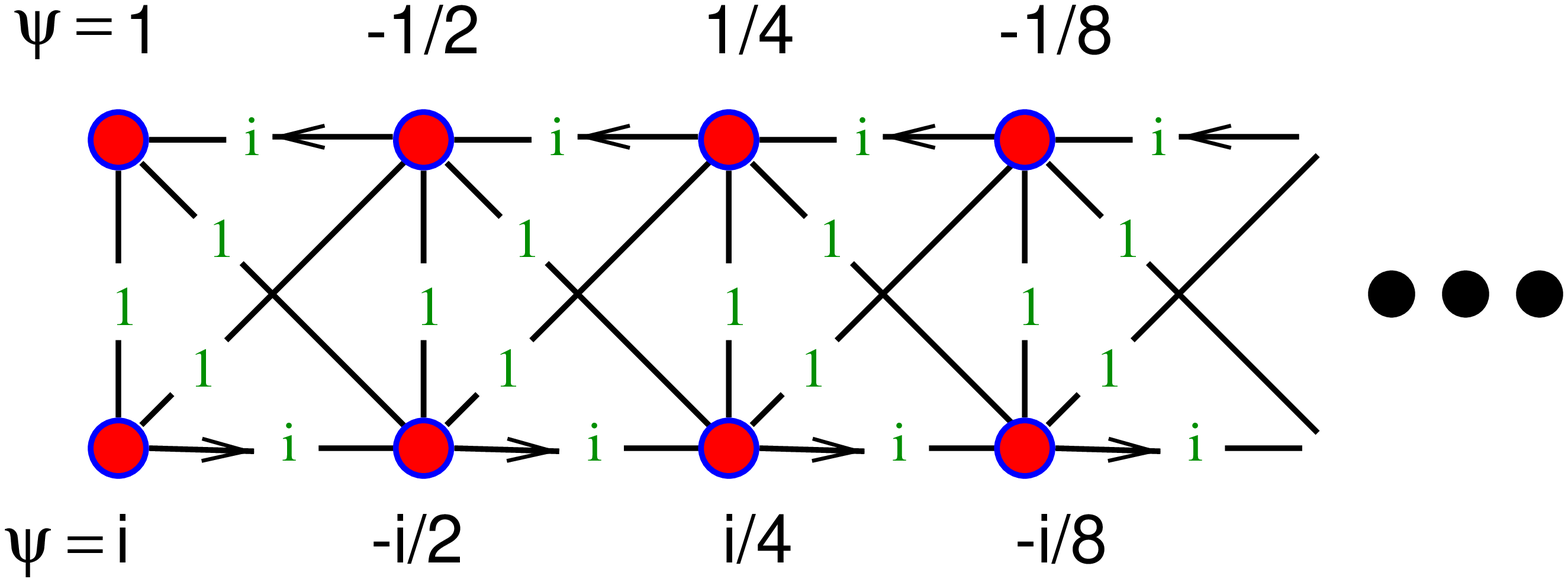}}
\medskip

When the chain is very long, both of the end states are forced to zero
energy by symmetry considerations.  First, since nothing distinguishes
one end of the chain from the other, they must have equal energy,
$E_L=E_R$.  On the other hand, a change in phase conventions,
effectively a gauge change, can change the sign of all the vertical
and diagonal bonds.  Following this with a left right flip of the
molecule will change the signs of the horizontal bonds.  This takes
the Hamiltonian to its negative, and shows that the states must have
opposite energies, $E_L=-E_R$.  This is indicative of a particle-hole
symmetry.  The combination of these results forces the end states to
zero energy, with no fine tuning of parameters.

For a finite chain, the exponentially decreasing penetration of the
end states into the molecule induces a small interaction between them.
They mix slightly to acquire exponentially small energies $E\sim \pm
e^{-\alpha L}$.  As the strength of the vertical bonds increases, so
does the penetration of the end states.  At a critical strength, the
mixing becomes sufficient that the zero modes blend into the positive
and negative energy bands.  In the full model, the mixing depends on
the physical momentum, and this disappearance of the zero modes is the
mechanism that removes the ``doublers'' when spatial momentum
components are near $\pi$ in lattice units \cite{mcih}.

Energy levels forced to zero by symmetry lie at the core of the domain
wall fermion idea.  On every spatial site of a three dimensional
lattice we place one of these chain molecules.  The distance along the
chain is usually referred to as a fictitious ``fifth'' dimension.  The
different spatial sites are coupled, allowing particles in the zero
modes to move around.  These are the physical fermions.  The
symmetries that protect the zero modes now protect the masses of these
particles.  Their masses receive no additive renormalization, exactly
the consequence of chiral symmetry in the continuum.  The physical
picture is sketched in this cartoon

\medskip
\epsfxsize .5\hsize
\centerline {\epsfbox{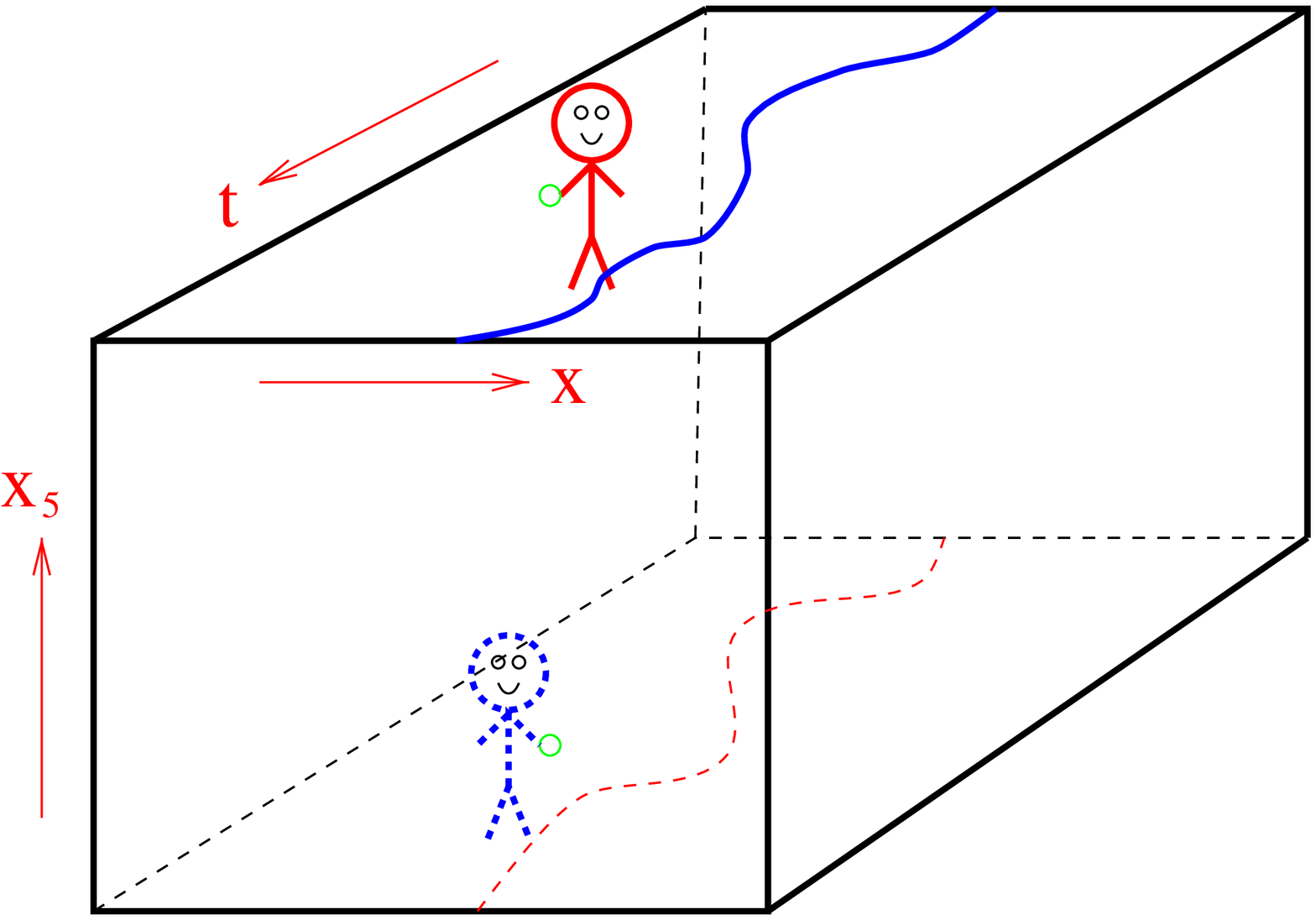}}

\noindent where I have rotated the fifth dimension to the vertical.
Our world lines traverse the four dimensional surface of this five
dimensional manifold.

Actually, the connection with chiral symmetry is much deeper than just
an analogy.  The construction guarantees that the modes are are
automatically chiral.  To see how this works, place Pauli spin
matrices on the spatial bonds.  This couples the phases seen by the
particles to their spins.  The zero mode that is attracted to one end
of the chain will continue to move spatially in a direction
corresponding to its helicity, as sketched here

\medskip
\epsfxsize .8\hsize
\centerline {\epsfbox{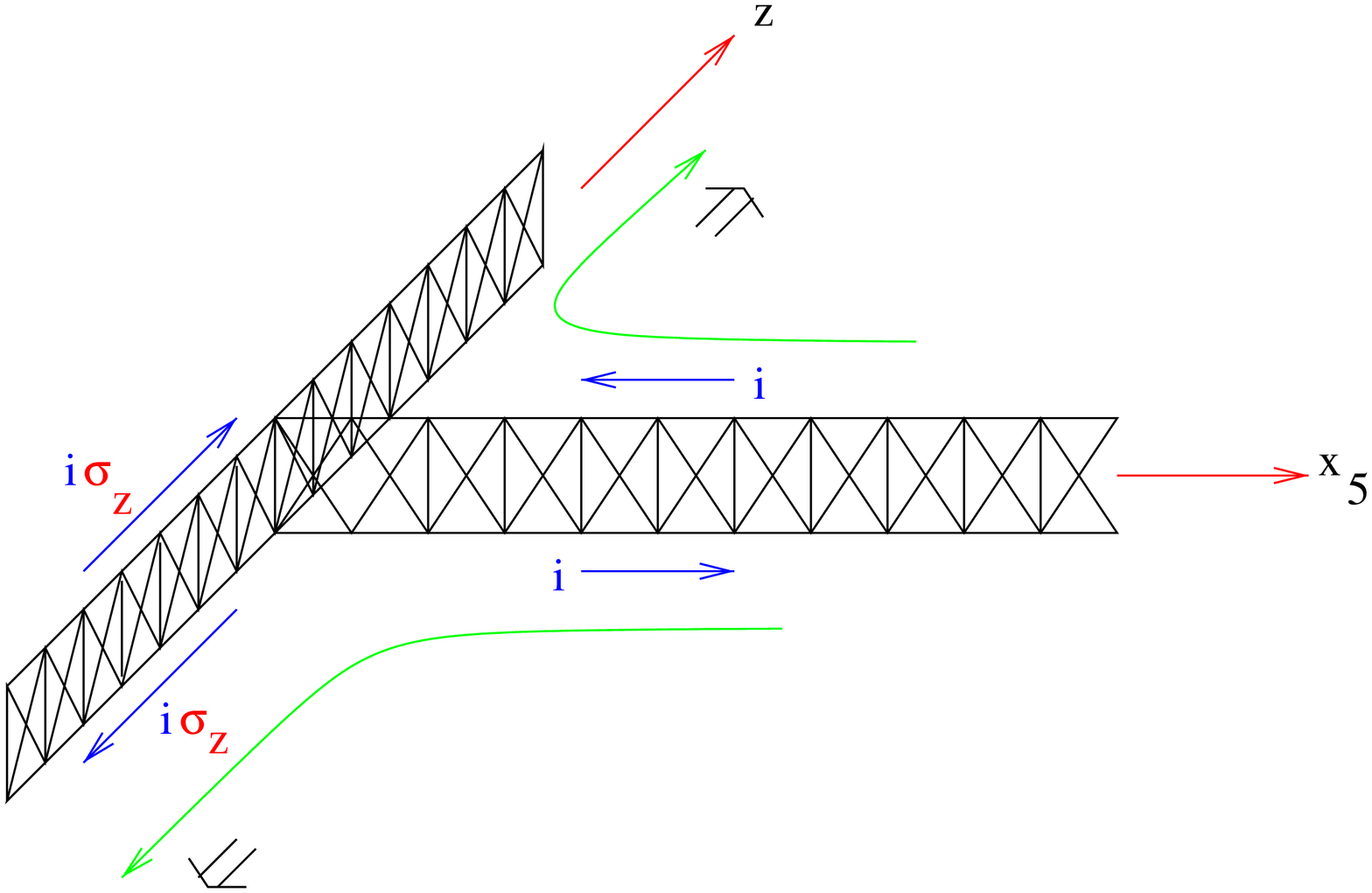}}

\noindent The ``device'' in this figure is effectively a
helicity projector.  This construction is equivalent to choosing a
particular set of Dirac gamma matrices
$$\matrix{
\gamma_0=\pmatrix{0 & 1 \cr 1 & 0\cr}\cr
\gamma_5=\pmatrix{0 & -i \cr i & 0\cr}\cr
\gamma_i=\pmatrix{0 & -i\sigma_i \cr i\sigma_i & 0\cr}
}
$$

\section{Slicing the fifth dimension}

I hope this description of domain-wall fermions in terms of simple
chain molecules has at least been thought provoking.  I now ramble on
with some general remarks about the basic scheme.  The existence of
the end states relies on using open boundary conditions in the fifth
direction.  If we were to curl our extra dimension into a circle, they
will be lost.  To retrieve them, consider cutting such a circle, as in
this figure

\epsfxsize .3\hsize
\centerline {\epsfbox{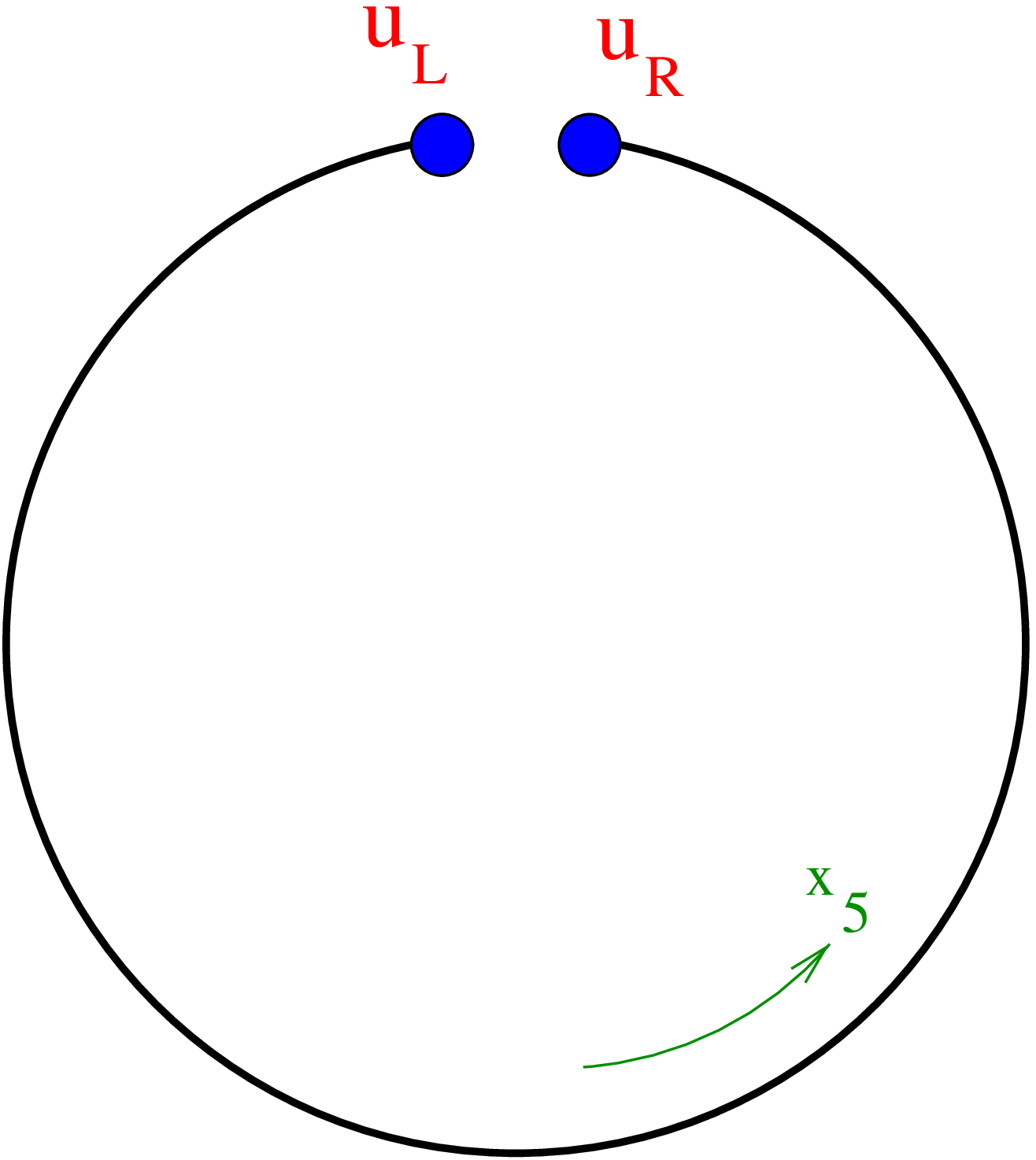}}

\noindent Of course, if the size of the extra dimension is finite, the
modes mix slightly.  This is crucial for the scheme to accommodate
anomalies \cite{mcih}.

Suppose I want a theory with two flavors of light fermion, such as the
up and down quarks.  For this one might cut the circle twice, as shown
here

\epsfxsize .3\hsize
\centerline {\epsfbox{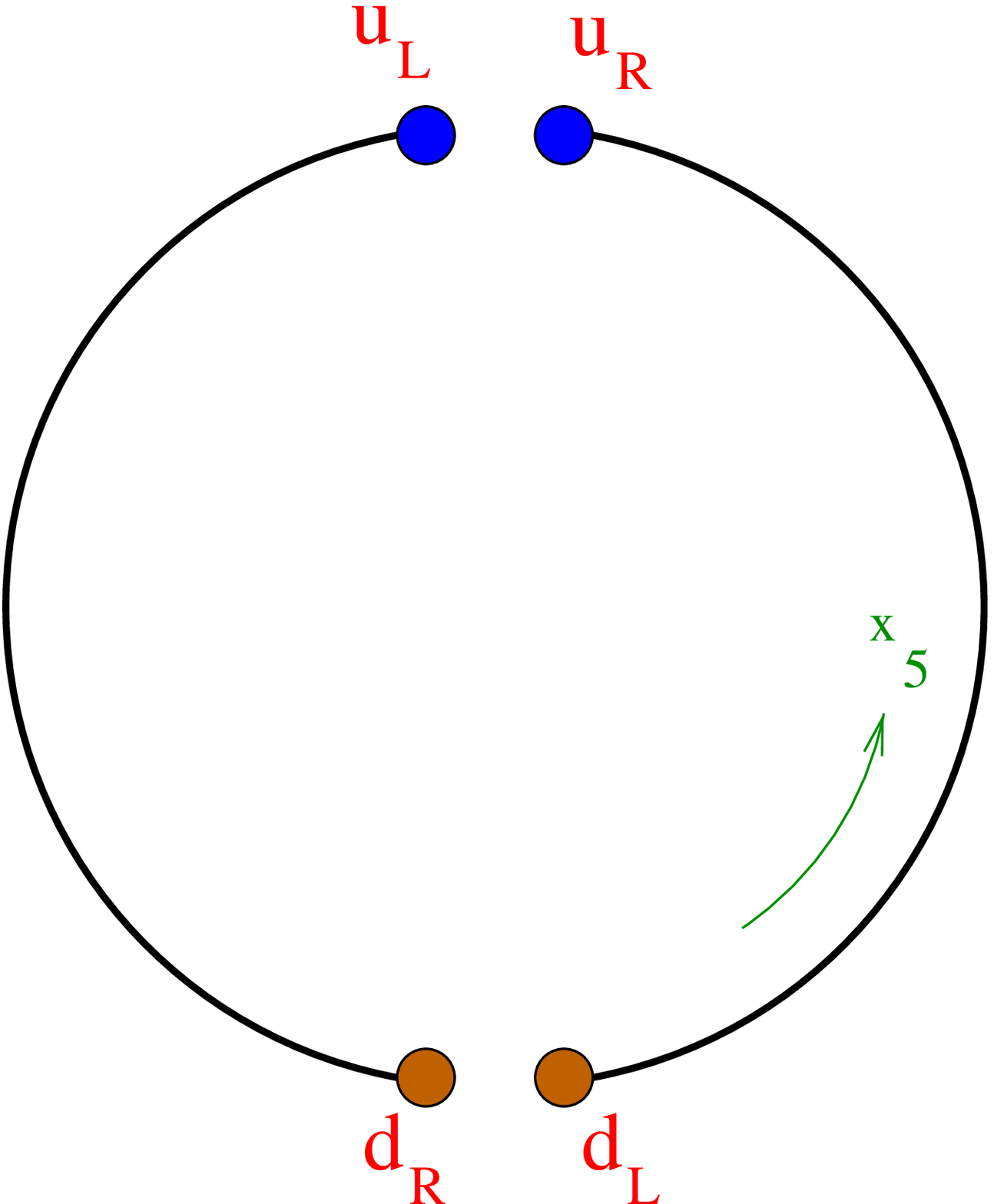}}
\noindent Remarkably, this construction keeps one chiral symmetry
exact, even if the size of the fifth dimension is finite.  Since the
cutting divides the molecule into two completely disconnected pieces,
in the notation of the figure we have the number of $u_L+d_R$
particles absolutely conserved.  Similarly with $u_R+d_L$.
Subtracting, we discover an exactly conserved axial charge
corresponding to the continuum current
$$
j_{\mu 5}^3 = \overline \psi \gamma_\mu\gamma_5 \tau^3 \psi
$$
The conservation holds even with finite $L_5$.  There is a small
flavor breaking since the $u_L$ mixes with the $d_R$.  These
symmetries are reminiscent of Kogut-Susskind \cite{kogutsusskind}, or
staggered, fermions, where a single exact chiral symmetry is
accompanied by a small flavor breaking.  Now, however, the extra
dimension gives additional control over the latter.

Despite this analogy, the situation is physically somewhat different
in the zero applied mass limit.  Staggered fermions are expected to
give rise to a single zero mass Goldstone pion, with the other pions
acquiring mass through the flavor breaking terms.  In my double cut
domain-wall picture, however, the zero mass limit has three degenerate
equal mass particles as the lowest states.  To see how this works it
is simplest to discuss the physics in a chiral Lagrangian language.
The finite fifth dimension generates an effective mass term, but it is
not in a flavor singlet direction.  Indeed, it is in a flavor
direction orthogonal to the naive applied mass.  In the usual
``sombrero'' picture of the effective Lagrangian, as illustrated here,

\bigskip
\epsfxsize .5\hsize \centerline {\epsfbox{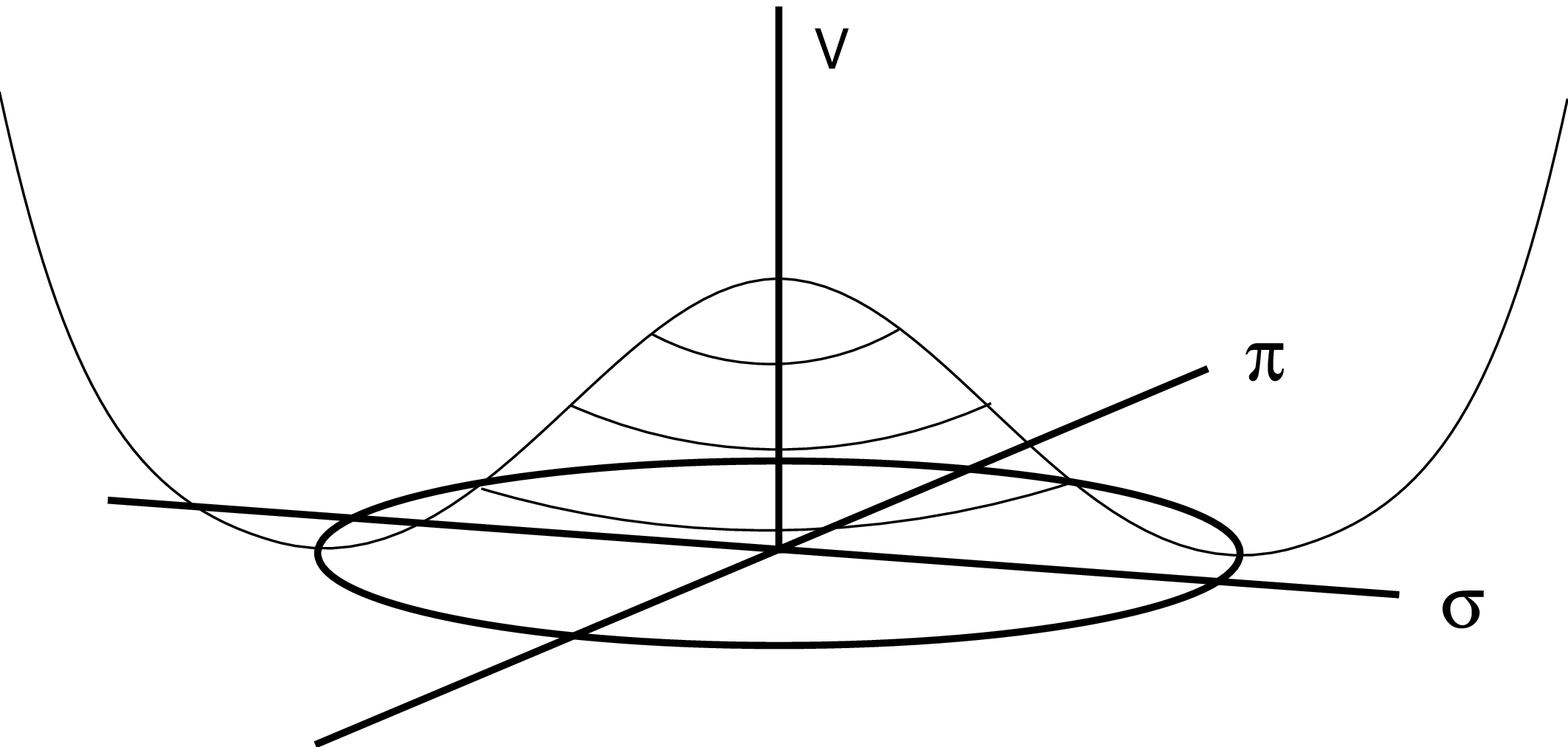}} 
\medskip
\noindent the two mass terms compete and the true vacuum rotates
around the Mexican hat from the conventional ``sigma'' direction to
the ``pi'' direction.

\section{How many fermions?}

Now I become more speculative.  The idea of cutting multiply the fifth
dimension to obtain several species suggests extensions to zero modes
on more complicated manifolds.  By having multiple zero modes, we have
a mechanism to generate multiple flavors.  Maybe one can have a theory
where all the physical fermions in four dimensions arise from a single
fermion field in the underlying higher dimensional theory.
Schematically we might have something like

\medskip
\epsfxsize .4\hsize
\centerline {\epsfbox{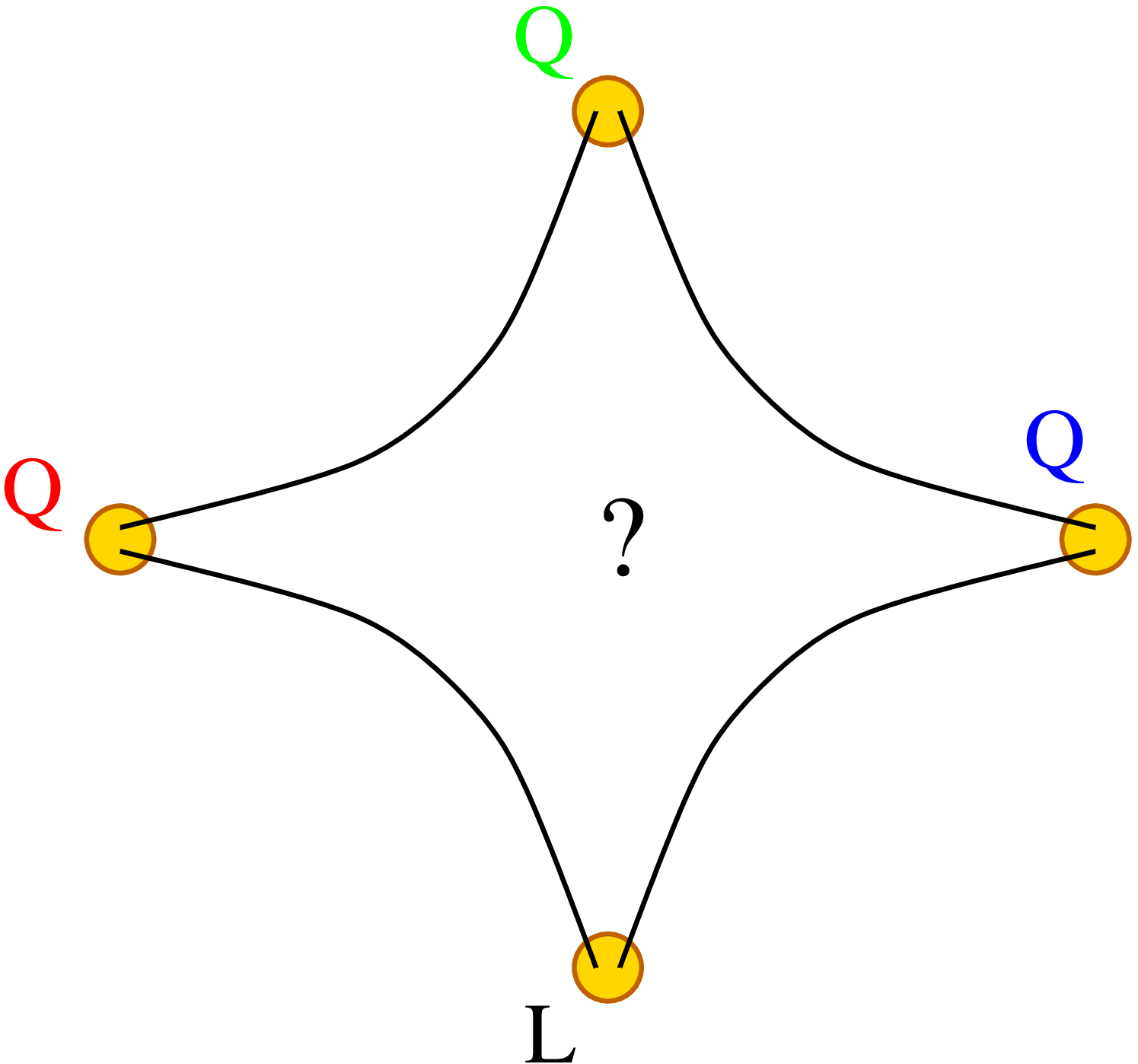}}
\noindent where each point represents some four dimensional surface
and the question remark represents structures in the higher dimension
that need specification.

One nice feature provided by such a scheme is a possible mechanism for
the transfer of various quantum numbers involved in anomalous
processes.  For example, the baryon non-conserving 't Hooft
process\cite{thooft} might arise from a lepton flavor tunneling into
the higher manifold and reappearing on another surface as a baryon.

I've been rather abstract here.  This generic mechanism is in fact the
basis of one specific proposed formulation of the standard model on
the lattice\cite{smol}.  For this model the question mark in the above
figure is a four-fermi interaction in the interior of the extra
dimension, as sketched here

\bigskip
\epsfxsize .6\hsize \centerline {\epsfbox{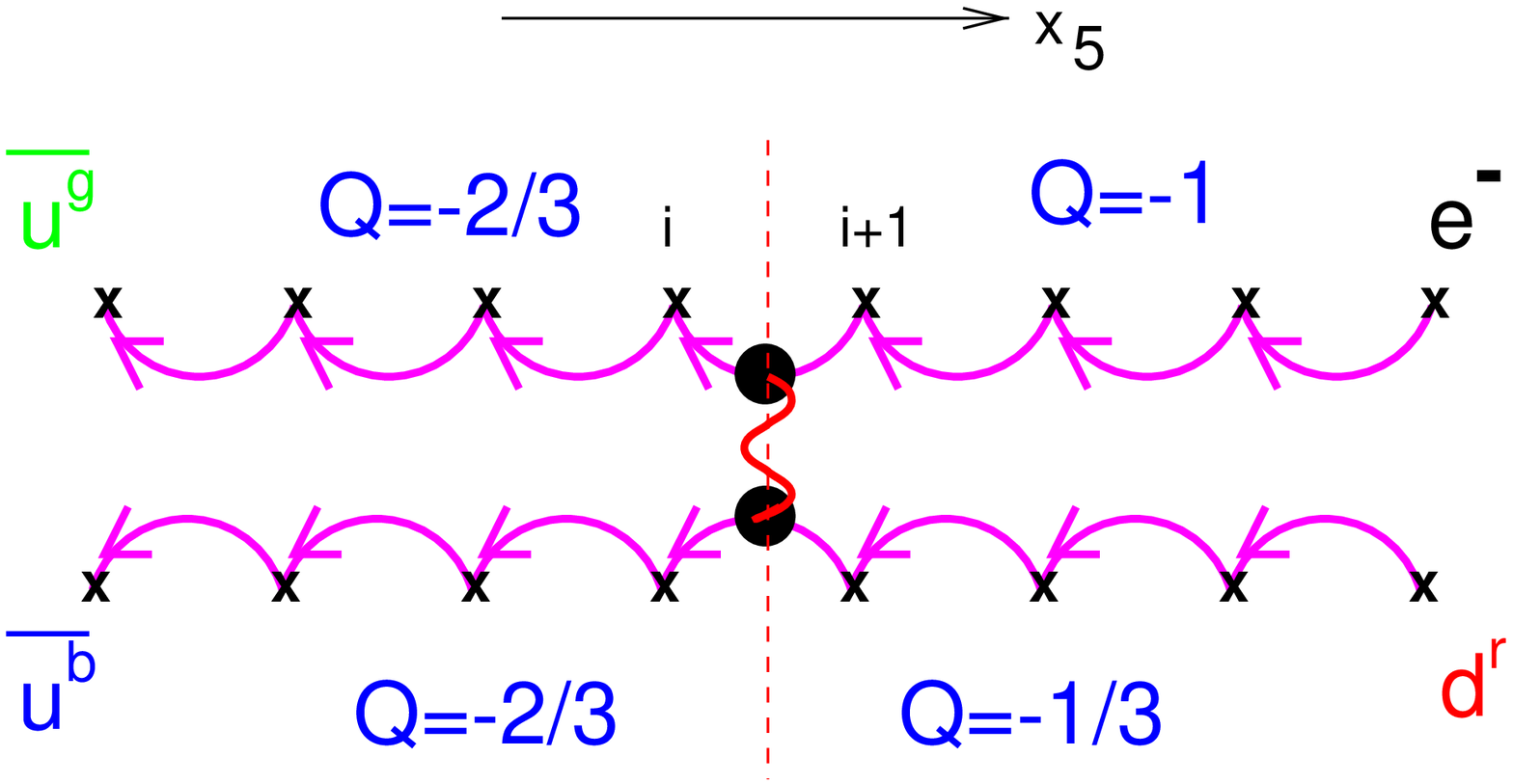}} 

\noindent In some sense, the right-handed doubler of the left-handed
electron is interpreted as an anti-quark.  This picture appears to
have all the necessary ingredients for a fully regulated and exactly
gauge invariant formulation of the standard model.  The primary
uncertainty lies with side effects of the multiple fermion coupling.
Our experience with non-perturbative phenomena involving strongly
coupled fermions is rather limited; in particular, the model cannot
tolerate the generation of a spontaneous symmetry breaking of any of
the gauge symmetries at the scale of the cutoff.

\section{Summary}

I have presented a simple molecular picture for zero modes protected
by symmetry.  This illustrates the mechanism for mass protection in
the domain-wall formulation of lattice fermions.  Then I discussed how
some chiral symmetries can become exact in this approach.  Finally I
speculated on schemes for generating multiple fermion species from the
geometry of higher dimensional models.  The latter may have
connections with the activities in string theory.

\section*{Acknowledgment}
This manuscript has been authored under contract number
DE-AC02-98CH10886 with the U.S.~Department of Energy.  Accordingly,
the U.S. Government retains a non-exclusive, royalty-free license to
publish or reproduce the published form of this contribution, or allow
others to do so, for U.S.~Government purposes.

\end{document}